\documentclass[preprint,review,12pt]{elsarticle}

\usepackage{graphicx}

\usepackage{amssymb}

\journal{Nuclear Instruments and Methods in Physics Research B }

\begin{document} 

\begin{frontmatter}



\title{Magnetic properties changes of MnAs thin films irradiated with highly charged ions}


\author[INSP]{M.~Trassinelli}
\ead{martino.trassinelli@insp.jussieu.fr}
\author[INSP]{V.E.~Gafton}
\author[INSP]{M.~Eddrief}
\author[INSP]{V.H.~Etgens}
\author[INSP]{E.~Lacaze}
\author[INSP]{E.~Lamour}
\author[INSP]{X.~Luo}
\author[INSP]{S.~Hidki}
\author[INSP]{M.~Marangolo}
\author[INSP]{J.~M\'erot}
\author[INSP]{C.~Prigent}
\author[INSP]{R.~Reuschl}
\author[INSP]{J.-P.~Rozet}
\author[INSP]{S.~Steydli}
\author[INSP]{D.~Vernhet}

\address[INSP]{Institut des NanoSciences de Paris, CNRS, Sorbonne Universit\'e -- Pierre et Marie Curie, UMR7588, 4 Place Jussieu, 75005 Paris, France}

\begin{abstract}
We present the first investigation on the effect of highly charged ion bombardment on a manganese arsenide thin film. 
The MnAs films, 150~nm thick, are irradiated with 90 keV Ne$^{9+}$ ions with a dose varying from $1.6\times10^{12}$ to $1.6\times10^{15}$~ions/cm$^2$. 
The structural and magnetic properties of the film after irradiation are investigated using different techniques, namely, X-ray diffraction, magneto-optic Kerr effect and magnetic force microscope.
Preliminary results are presented.
From the study of the lattice spacing, we measure a change on the film structure that depends on the received dose, similarly to previous studies with other materials. 
Investigations on the surface show a strong modification of its magnetic properties.

\end{abstract}


\end{frontmatter}


\section{Introduction}
\label{sec:intro}
Thin films of ferromagnetic transition metals have been extensively studied in the last decades for their important applications.
Many investigations on the modification of their magnetic properties have been performed using different techniques. In particular, ion bombardment demonstrated to be an efficient tool for this task \cite{Chappert1998}.
So far, experiments were performed on transition ferromagnetic metals and on some of their corresponding alloys.
They demonstrated that ion bombardment does not affect the chemical composition of the films, but changes significantly structure and magnetic properties. 
These modifications are related to local changes of the film structure during the irradiation and to the implantation of ions.
More precisely,  dependencies on the ion specie, its penetration depth and irradiation dose, giving rises to changes on the coercivity, saturation magnetization and crystal lattice, have been observed \cite{Kaminsky2001,Zhang2003,Zhang2004,Muller2005,Fassbender2006,Cook2011}.

In the this work, we present recent studies on more peculiar magnetic material, namely the manganese arsenide, submitted to ion bombardment.
MnAs is a metallic compound that is ferromagnetic below T$_C = 313$~K where the first-order phase transition from hexagonal ($\alpha$ phase) to orthorhombic ($\beta$ phase, MnP type) is accompanied by a ferromagnetic-paramagnetic transition.
The possibility of epitaxial growth of MnAs thin films on standard semiconductors such as GaAs has renewed interest for spintronic research \cite{Daweritz2006}. 
The epitaxial strain disturbs the phase transition that leads to the $\alpha - \beta$ phase coexistence over a large range of temperatures (10--45$^\circ$C \cite{Kaganer2002}).
In this range, an alternating structure of ridges ($\alpha$ phase) and grooves ($\beta$ phase) organized in stripe-shaped domains is created to minimize the elastic energy due to the epitaxy.

Very recently, our group has started to study the change of these peculiar characteristics  induced by the implantation of Ne$^{9+}$ highly charged ions in the MnAs epilayer. 
Such changes may open new perspectives in the attractive possibility to maintain ferromagnetism in MnAs to higher temperature than the bulk T$_C$.
A better knowledge of these damages will also provide important information about the robustness of this material in hostile radiation environment.  

Production of the samples, irradiation and characterization have been entirely performed at the NanoScience Institute of Paris (INSP) where a molecular beam epitaxy (MBE) system and an ion source equipped with a dedicated beam line (SIMPA, French acronym for highly charged ion source of PAris)  are available. 
After the irradiation, the sample is characterized by several techniques: MOKE for magneto-optical Kerr effect, XRD for  X-ray diffractometer and MFM for  magnetic force microscope.  
In this paper, we detail the production of MnAs/GaAs films and the set-up for ion irradiation. 
First results and discussions of the bombarded films are presented.

\section{Experimental details}

\subsection{MnAs film production}
MnAs epilayers are grown by MBE on GaAs(001) substrate. The MnAs growth is performed at 240$^\circ$C on a prepared GaAs substrate under As-rich conditions. A growth rate of about 100nm/h leads to a finally thickness of 150~nm. 
The deposed MnAs is oriented with the $\alpha$MnAs $[0001]$ and $\beta$MnAs $[001]$ axis parallel to GaAs$[\bar110]$.
At the end of the growing processes the samples are capped in situ with an amorphous As layer in order to prevent the MnAs oxidation. 
More details on the growing process can be found in \cite{Breitwieser2009}.

Samples  of same MnAs/GaAs growth with a surface of $4 \times 5$~mm$^2$ are obtained from the original wafer.  
Once placed on a Omicron sample holder, the films are moved toward the SIMPA facility in atmosphere pressure.

\subsection{Ion irradiation}
The ion irradiation are performed at the SIMPA facility that include an electron-cyclotron resonance ion source coupled to a dedicated beam line. 
The ion source can produce intense beams of highly charged ions such as Ne$^{9+}$, Ar$^{16+}$, Xe$^{26+}$ \ldots \
After the extraction, the ion beam is selected in charge and mass by a auto-focusing magnetic dipole. With a series of magnetic and electric lenses, the ion beam is transported to the collision chamber where the different samples are placed. 
In our particular case, we use a beam  Ne$^{9+}$ with a kinetic energy of the ions of 90~keV (4.5~keV/u) and a maximum intensity of 1~e$\mu$A in the chamber. 
More details of the SIMPA installation can be found in ref. \cite{Gumberidze2010}.

\begin{figure}[ht]
\begin{center}
\includegraphics[trim=2.cm 1cm 2.cm 1cm, clip=true,width=\textwidth]{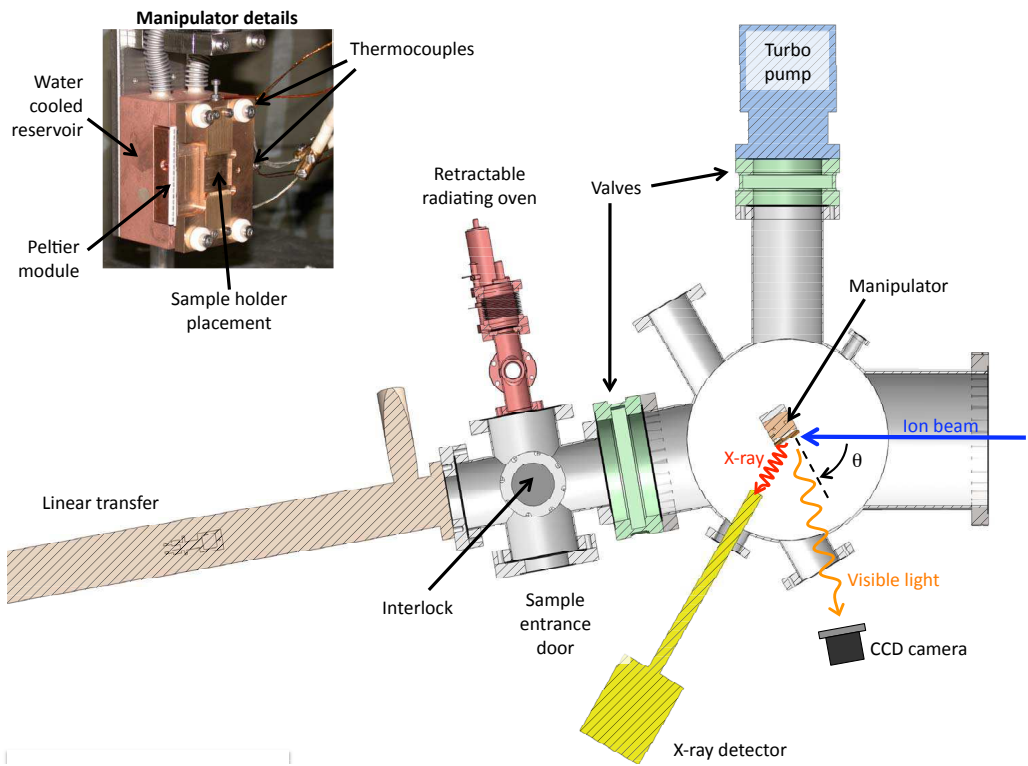}
\caption{Scheme of the collision chamber. The interlock and the different diagnostic tools are shown. In the inset, a picture of the sample manipulator, thermally regulated, is presented.}
\label{fig:chamber}
\end{center}
\end{figure}

The samples are introduced in the collision chamber via an interlock equipped with a radiative oven (see figure \ref{fig:chamber}).
In the interlock, once a pressure of few 10$^{-8}$~mbar is reached, each sample is slowly warmed up to 270$^\circ$C and kept to this temperature for several hours in order to remove the As cap. 
Once cooled back to room temperature, the sample is transferred to the sample-manipulator located at the center of the collision chamber.
Typical pressure of the chamber is $5 \times 10^{-9}$~mbar.
The manipulator allows for xyz translation and a rotation long the axis perpendicular to the ion beam to regulate the incidence angle of the ions on the sample.
During the irradiation, the sample temperature can be controlled  trough  a series of K-type thermocouples and eventually changed with a thermal regulator composed by a Peltier cooler between the sample holder and a water cooled component.

\begin{figure}
\begin{center}
\includegraphics[width=0.45\textwidth]{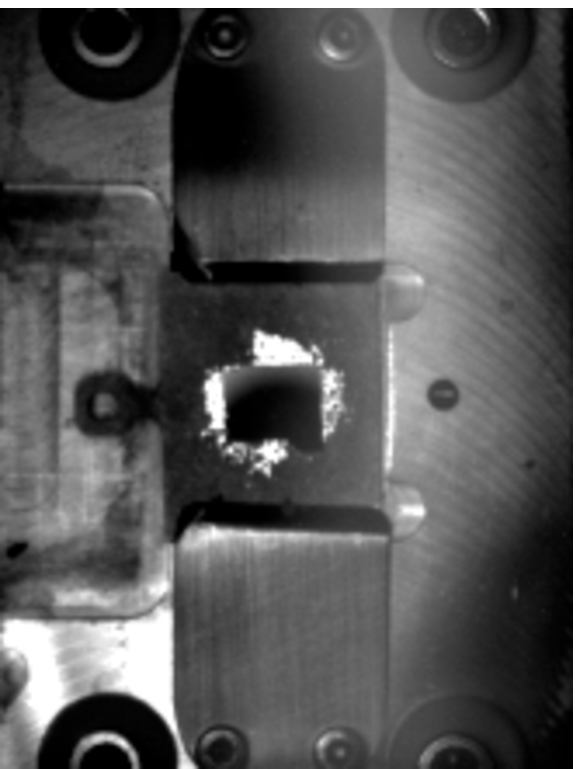} \hfill
\includegraphics[width=0.45\textwidth]{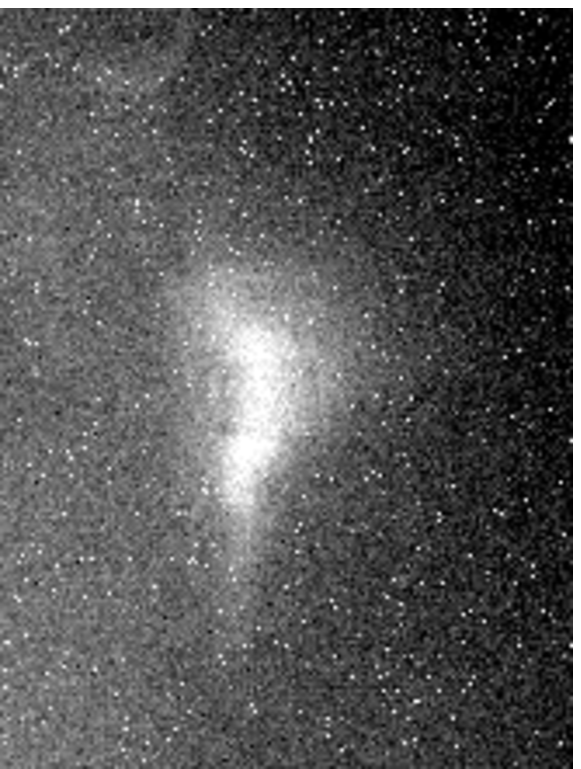}
\caption{Left: image of the sample, illuminated by external light on a Omicron sample holder mounted on the manipulator in the center of the collision chamber. Right: the same sample under ion bombardment. The shape and position of the beam hitting the sample is well visible.}
\label{fig:sample}
\end{center}
\end{figure}

The ion beam profile and position are regulated and controlled by means of a high sensitive Charged Coupe Device (CCD) and adjustable magnetic steerers and electrostatic lenses before the entrance of the chamber. 
The CCD, placed at an observing angle of about 80$^\circ$, images the ion collision florescence light produced either on the stainless steel surface mounted on the back of the sample-manipulator or on the sample itself.
The ion beam current is measured with a removable Faraday cup in front of the sample. 
The incidence angle between the ion beam and the sample surface is fixed to 60$^\circ$. This choice of angle enables to monitor the ion--sample collision region with the CCD camera and with an X-ray solid state detector during the irradiation time.
The detection of the characteristic Ne$^{9+}$ X-ray emission produced during the interaction of the ions with the sample provides, in fact, an additional control on the ion beam intensity during the irradiation.
Typical pictures of the sample illuminated from external light and under ion bombardment are presented in figure~\ref{fig:sample}.

\begin{figure}
\begin{center}

\includegraphics[width=0.47\textwidth]{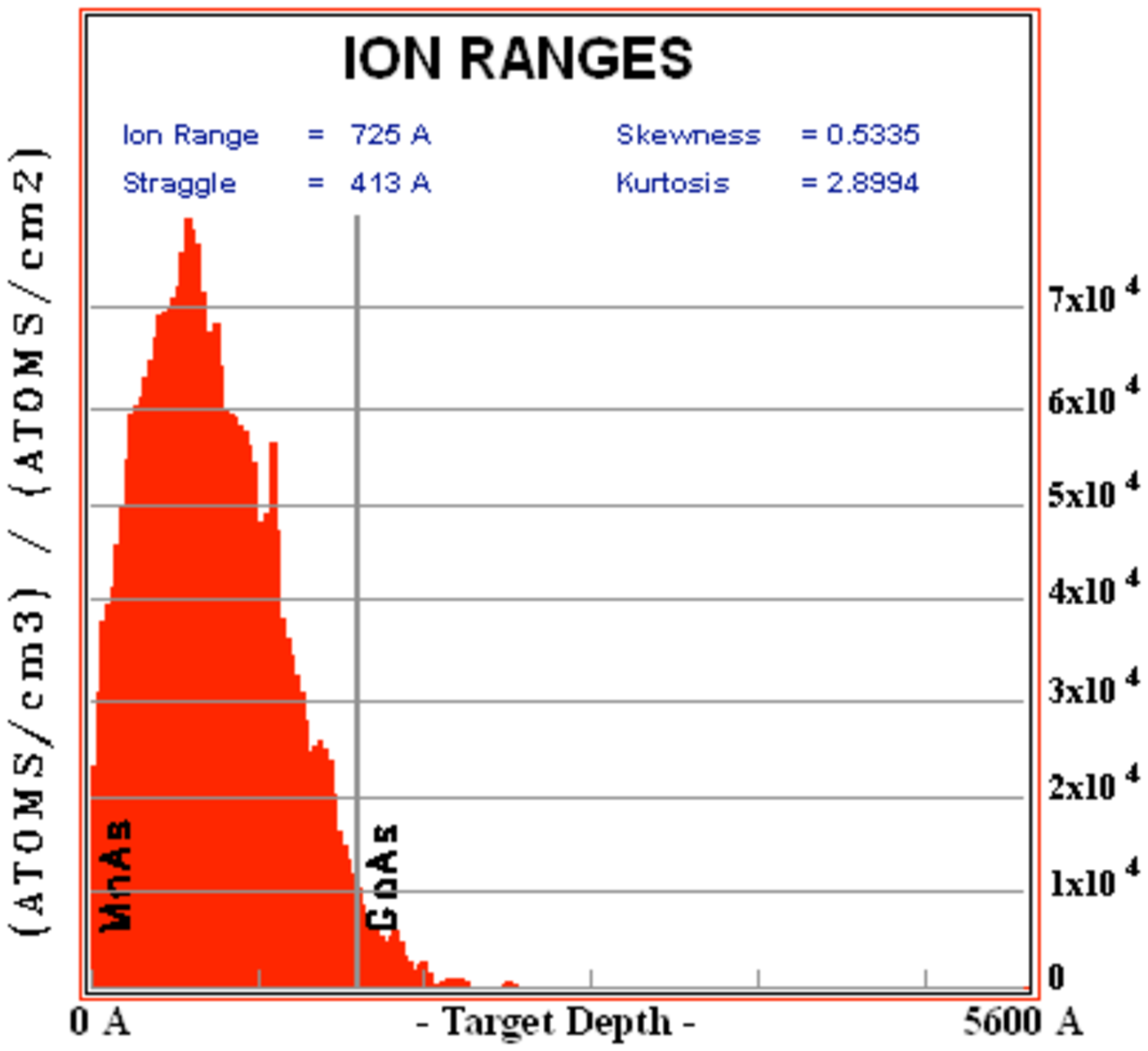} \hfill
\includegraphics[width=0.47\textwidth]{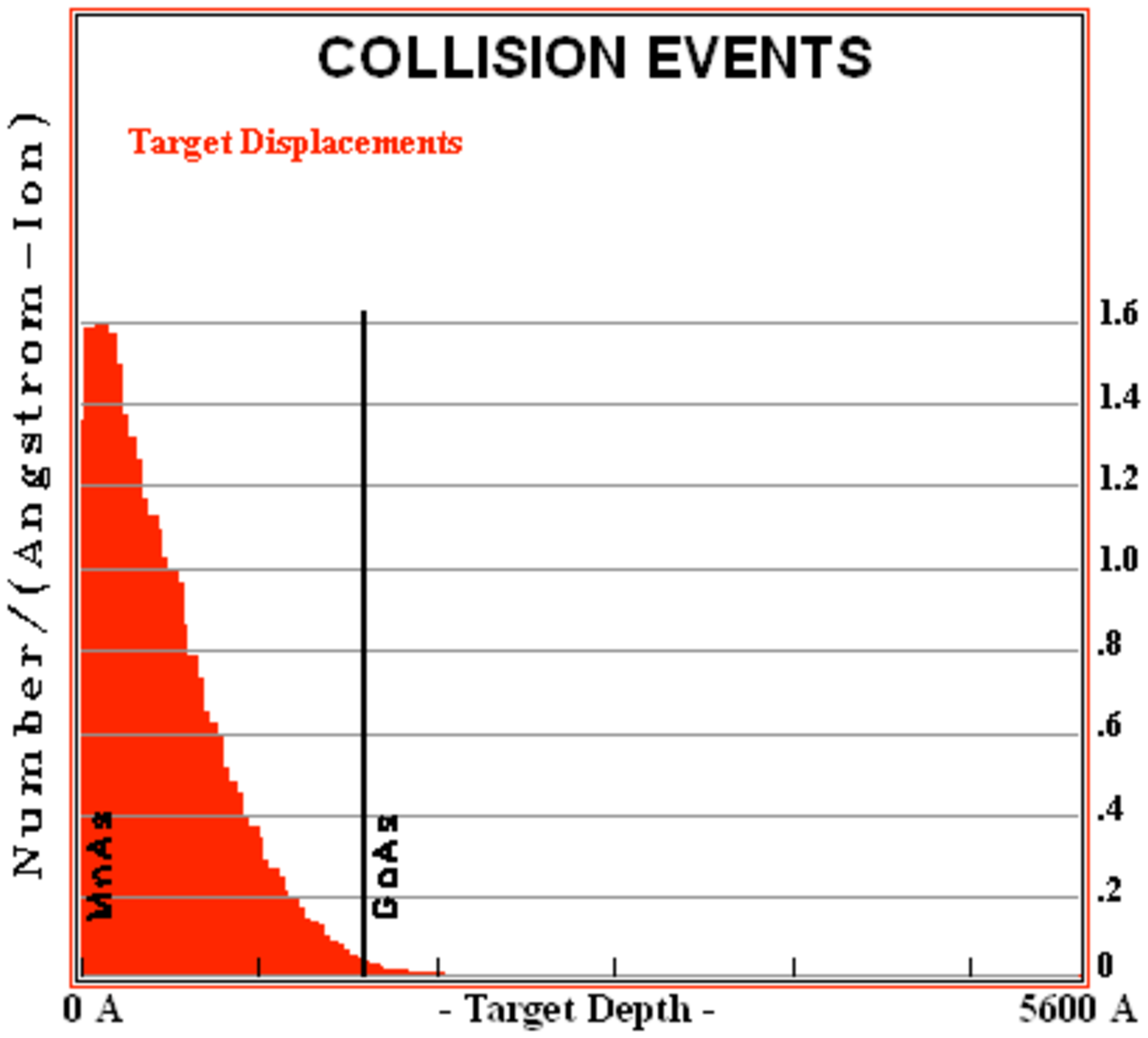}
\caption{SRIM predictions \cite{Ziegler2008} on the ion penetration and collision events with the present experimental conditions: Ne$^{9+}$ at 90~keV on MnAs/GaAs (150 nm) with an incidence angle of 60$^\circ$. Left: distribution of the ion penetration. Right: distribution of atom displacement in the sample. As we can observe, only a very small fraction of ions is deposited in the GaAs substrate.}
\label{fig:SRIM}
\end{center}
\end{figure}

Under these experimental condition, we determine from SRIM (version 2008.04 \cite{Ziegler2008}) that the ion implanted distribution covers the full MnAs film thickness enhancing hence the dose effect \cite{Zhang2004}.
The simulations in figure~\ref{fig:SRIM} show that only a very small fraction of ions is deposited in the GaAs substrate excluding the possibility of MnAs-GaAs beam mixing.

Different ion beam bombardment durations, from 5 to several thousands of seconds, corresponding to doses between  $1.6\times10^{12}$ and  $1.6\times10^{15}$~ions/cm$^2$, are applied on different samples coming from the same growth. 
Due to the difficulty to have a well defined beam shape on the sample, the dose is estimated with a poor accuracy evaluated to about 50\%.

\section{Analysis of the irradiated samples and discussions}
After the irradiation, the samples are removed from the interaction chamber to be analyzed.
Preliminary studies with MOKE techniques that  measure the macroscopic magnetic response at the surface. This first results indicate that already with $1.6\times10^{12}$~ions/cm$^2$ no ferromagnetic behavior is observed.
More systematic analysis are performed using XRD and MFM methods and they are presented in the following paragraphs.

\subsection{XRD measurements}
The acquisition of XRD data is performed using a PANalytical XPertPRO MRD diffractometer with a copper anode, generating CuK$\alpha_{1,2}$ radiation and operating at 40~kV and 30~mA. The diffractometer is equipped with 4 circular cradles to move the sample with respect different angles for positioning and scanning. XRD patterns are recorded in the $2 \theta$ range $20-100^\circ$, with a step of $0.05^\circ$, and a 17~s time per step. 
With XRD measurements at grazing incidence angle, the thickness of the sample is determined before and after irradiation. 
All measurements are obtained at a temperature of $20\pm1^\circ$C, where $\alpha$ and $\beta$ phases are normally present in non irradiated MnAs/GaAs films.
Data from irradiated films are compared with a reference decapped MnAs/GaAs sample issued from the same growth.

\begin{figure}[ht!]
\begin{center}

\includegraphics[width=0.7\textwidth]{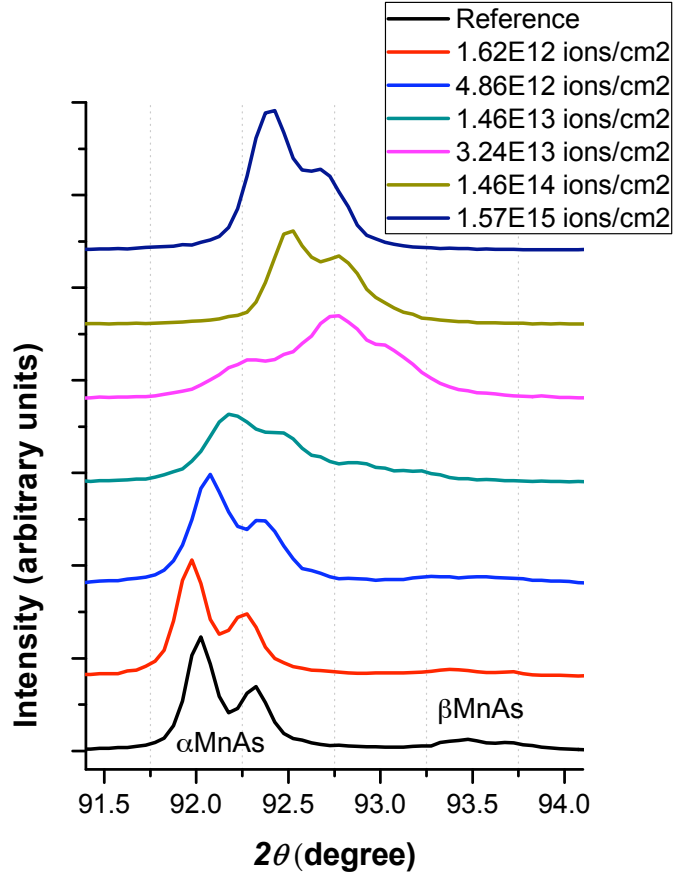}
\caption{XRD Bragg peaks of $\alpha$MnAs$(\bar3300)$ and $\beta$MnAs$(060)$ plans as a function of the received doses (in ions per cm$^2$).}
\label{fig:XRD}
\end{center}
\end{figure}

The thickness of the samples is found to remain constant whatever the ion dose (within the range used here) with a value around 150~nm demonstrating that sputtering effects are marginal.
In opposite, the Bragg reflection spectra exhibits significant modifications as a function of the received dose, as seen in figure~\ref{fig:XRD}.
For the non irradiated sample, the $\alpha$MnAs$(\bar3300)$ and $\beta$MnAs$(060)$ reflection peaks are clearly identified (the $\beta$ peaks are much less intense at 20$^\circ$C). 
The double peak structure is due to the CuK$\alpha_{1,2}$ doublet used in the present spectrometer. 
For low doses, the $\beta$MnAs$(060)$ peak is still visible but seems to disappears for doses $> 5\times10^{12}$~ions/cm$^2$. For doses around $2\times10^{12}$~ions/cm$^2$, the $\alpha$ and $\beta$ peaks are strongly distorted ending to a unique doublet at higher doses ($\geq1.5\times10^{14}$~ions/cm$^2$).
This Bragg reflection could correspond to a new phase of MnAs with Ne implantation or to a modification of the $\alpha$ or $\beta$ phases in analogy of the measurement on Co films bombarded with Xe ions reported in \cite{Zhang2003}.
In this previous work, local warming during the ion impact induces a phase transition between the hexagonal close packing (hcp) phase, stable in bulk Co below 425$^\circ$C, and the face-centered cubic (fcc) phase, stable at higher temperature.
An analog phenomenon could appear in our case between the two MnAs phases.  
To probe this aspect, additional measurements on magnetic properties are required to detect the presence or not of the ferromagnetic $\alpha$MnAs phase.
As mentioned before, MOKE measurements suggest that MnAs after Ne bombardment has no macroscopic magnetic properties.
For a closer investigation on the local magnetic properties on the surface, we used an MFM. 
The results are presented in the following section.

\subsection{MFM measurements}
The magnetic images of the sample surface are taken with a VEECO dimension 3100 MFM equipped with a magnetic tip coated with Co/Cr (model MESP) and working at a temperature around 20$^\circ$C. 
Before each scan, the samples are magnetized long the surface in parallel to the easy axis of the MnAs $[\bar1100]$. 

\begin{figure}
\begin{center}

\includegraphics[width=\textwidth]{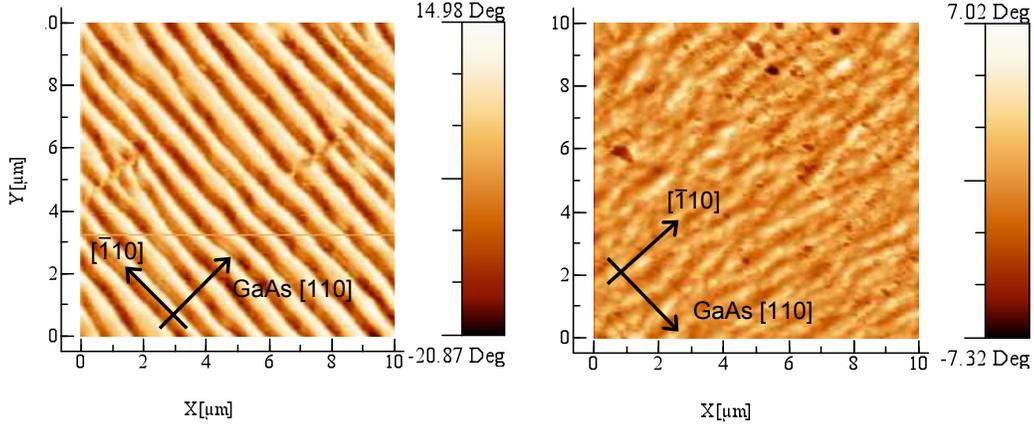}
\caption{MFM images of 150 nm MnAs on GaAs before (left) and after (right) ion bombardment with a dose of about $1.5\times10^{13}$ ions/cm$^2$. The orientation axis of the GaAs substrates are indicated on each image.}
\label{fig:MFM}
\end{center}
\end{figure}

MFM images of the decapped samples before irradiation show (see figure~\ref{fig:MFM} left as an example) the well defined and characteristic stripes of MnAs on GaAs due to the succession of $\alpha$MnAs and $\beta$MnAs regions.
We observed that, after irradiation, the shape of the stripes is more and more distorted and at  $1.5\times10^{13}$ ions/cm$^2$ the stripes are completely vanished. Instead, chaotic pattern with elongated regions parallels to the original stripes (parallel to GaAs$[\bar110]$) appears, as it can be observed in figure~\ref{fig:MFM} (right). 
At higher irradiation doses, the phase contrast continues to decreases toward an almost complete disappearance.

\subsection{Discussions}
Due to the presence of the phase contrast of the MFM images, we can deduce that a local magnetic response is still present at relative at irradiation doses $\leq 3.2\times10^{13}$ ions/cm$^2$ but in a meantime, the $\alpha$ and $\beta$ phases are not anymore distinguishable.  
In contrast, MOKE data, sensitive to the macroscopic magnetic response of the surface, show that no signal of ferromagnetic behavior is detected with samples irradiated with a dose equal or lower than $1.2\times10^{12}$ ions/cm$^2$. 
This could indicate of a freezing of the magnetic domain alignment on the surface, even when an external field is applied (as with MOKE).
Considering also the XRD data, we can emphasize that these new properties could be intrinsically connected to the presence of a new phase.
To confirm or disprove this assertion, more investigation will be performed in the future.

\section{Conclusions}
First investigation on the effect of Ne$^{9+}$ highly charged ion bombardment on MnAs/GaAs thin film is presented. 
The structural and magnetic properties of the MnAs film before and after irradiation are investigated using different techniques, namely, X-ray diffraction, magneto-optic Kerr effect and magnetic force microscope.
Preliminary results indicated that ferromagnetic properties of irradiated samples are visible locally using MFM but strongly distorted when compared to non irradiated samples.
XRD measurements indicate that this behavior could be linked on a presence of a strongly modified $\alpha$ phase of MnAs or even a new phase.
More studies are in any case required.

\section*{Acknowledgments}
The authors would like to thank Mariana Barturen for her help on the MOKE measurements.
This experiment has been partially supported by a grant from 'Agence Nationale pour la Recherche (ANR)' number ANR-06-BLAN-0223.






\begin{thebibliography}{12}
\expandafter\ifx\csname natexlab\endcsname\relax\def\natexlab#1{#1}\fi
\providecommand{\bibinfo}[2]{#2}
\ifx\xfnm\relax \def\xfnm[#1]{\unskip,\space#1}\fi
\bibitem[{Chappert et~al.(1998)Chappert, Bernas, Ferr\'e, Kottler, Jamet, Chen,
  Cambril, Devolder, Rousseaux, Mathet, and Launois}]{Chappert1998}
\bibinfo{author}{C.~Chappert}, \bibinfo{author}{H.~Bernas},
  \bibinfo{author}{J.~Ferr\'e}, \bibinfo{author}{V.~Kottler},
  \bibinfo{author}{J.~P. Jamet}, \bibinfo{author}{Y.~Chen},
  \bibinfo{author}{E.~Cambril}, \bibinfo{author}{T.~Devolder},
  \bibinfo{author}{F.~Rousseaux}, \bibinfo{author}{V.~Mathet},
  \bibinfo{author}{H.~Launois},
\newblock \bibinfo{title}{Planar patterned magnetic media obtained by ion
  irradiation},
\newblock \bibinfo{journal}{Science} \bibinfo{volume}{280}
  (\bibinfo{year}{1998}) \bibinfo{pages}{1919--1922}.
\bibitem[{Kaminsky et~al.(2001)Kaminsky, Jones, Patel, Booij, Blamire,
  Gardiner, Xu, and Bland}]{Kaminsky2001}
\bibinfo{author}{W.~M. Kaminsky}, \bibinfo{author}{G.~A.~C. Jones},
  \bibinfo{author}{N.~K. Patel}, \bibinfo{author}{W.~E. Booij},
  \bibinfo{author}{M.~G. Blamire}, \bibinfo{author}{S.~M. Gardiner},
  \bibinfo{author}{Y.~B. Xu}, \bibinfo{author}{J.~A.~C. Bland},
\newblock \bibinfo{title}{Patterning ferromagnetism in {Ni$_{80}$Fe$_{20}$}
  films via {Ga}$^+$ ion irradiation},
\newblock \bibinfo{journal}{Appl. Phys. Lett.} \bibinfo{volume}{78}
  (\bibinfo{year}{2001}) \bibinfo{pages}{1589--1591}.
\bibitem[{Zhang et~al.(2003)Zhang, Gupta, Lieb, Luo, M\"uller, Schaaf, and
  Uhrmacher}]{Zhang2003}
\bibinfo{author}{K.~Zhang}, \bibinfo{author}{R.~Gupta}, \bibinfo{author}{K.~P.
  Lieb}, \bibinfo{author}{Y.~Luo}, \bibinfo{author}{G.~A. M\"uller},
  \bibinfo{author}{P.~Schaaf}, \bibinfo{author}{M.~Uhrmacher},
\newblock \bibinfo{title}{Xenon-ion--induced phase transition in thin {Co}
  films},
\newblock \bibinfo{journal}{Eur. Phys. Lett.} \bibinfo{volume}{64}
  (\bibinfo{year}{2003}) \bibinfo{pages}{668}.
\bibitem[{Zhang et~al.(2004)Zhang, Lieb, M\"uller, Schaaf, Uhrmacher, and
  M\"unzenberg}]{Zhang2004}
\bibinfo{author}{K.~Zhang}, \bibinfo{author}{K.~P. Lieb},
  \bibinfo{author}{G.~A. M\"uller}, \bibinfo{author}{P.~Schaaf},
  \bibinfo{author}{M.~Uhrmacher}, \bibinfo{author}{M.~M\"unzenberg},
\newblock \bibinfo{title}{Magnetic texturing of xenon-ion irradiated nickel
  films},
\newblock \bibinfo{journal}{Eur. Phys. J. B} \bibinfo{volume}{42}
  (\bibinfo{year}{2004}) \bibinfo{pages}{193--204}.
\bibitem[{M\"uller et~al.(2005)M\"uller, Carpene, Gupta, Schaaf, Zhang, and
  Lieb}]{Muller2005}
\bibinfo{author}{G.~A. M\"uller}, \bibinfo{author}{E.~Carpene},
  \bibinfo{author}{R.~Gupta}, \bibinfo{author}{P.~Schaaf},
  \bibinfo{author}{K.~Zhang}, \bibinfo{author}{K.~P. Lieb},
\newblock \bibinfo{title}{Ion-beam induced changes in magnetic and
  microstructural properties of thin iron films},
\newblock \bibinfo{journal}{Eur. Phys. J. B} \bibinfo{volume}{48}
  (\bibinfo{year}{2005}) \bibinfo{pages}{449--462}.
\bibitem[{Fassbender et~al.(2006)Fassbender, von Borany, M\"ucklich, Potzger,
  M\"oller, McCord, Schultz, and Mattheis}]{Fassbender2006}
\bibinfo{author}{J.~Fassbender}, \bibinfo{author}{J.~von Borany},
  \bibinfo{author}{A.~M\"ucklich}, \bibinfo{author}{K.~Potzger},
  \bibinfo{author}{W.~M\"oller}, \bibinfo{author}{J.~McCord},
  \bibinfo{author}{L.~Schultz}, \bibinfo{author}{R.~Mattheis},
\newblock \bibinfo{title}{Structural and magnetic modifications of
  {Cr}-implanted {Permalloy}},
\newblock \bibinfo{journal}{Phys. Rev. B} \bibinfo{volume}{73}
  (\bibinfo{year}{2006}) \bibinfo{pages}{184410}.
\bibitem[{Cook et~al.(2011)Cook, Shen, Grundy, Im, Fischer, Morton, and
  Kilcoyne}]{Cook2011}
\bibinfo{author}{P.~J. Cook}, \bibinfo{author}{T.~H. Shen},
  \bibinfo{author}{P.~J. Grundy}, \bibinfo{author}{M.~Y. Im},
  \bibinfo{author}{P.~Fischer}, \bibinfo{author}{S.~A. Morton},
  \bibinfo{author}{A.~L.~D. Kilcoyne},
\newblock \bibinfo{title}{Focused ion beam patterned {Fe} thin films: A study
  by selective area {Stokes} polarimetry and soft x-ray microscopy},
\newblock \bibinfo{journal}{J. Appl. Phys.} \bibinfo{volume}{109}
  (\bibinfo{year}{2011}) \bibinfo{pages}{063917--5}.
\bibitem[{D\"aweritz(2006)}]{Daweritz2006}
\bibinfo{author}{L.~D\"aweritz},
\newblock \bibinfo{title}{Interplay of stress and magnetic properties in
  epitaxial {MnAs} films},
\newblock \bibinfo{journal}{Rep. Prog. Phys.} \bibinfo{volume}{69}
  (\bibinfo{year}{2006}) \bibinfo{pages}{2581}.
\bibitem[{Kaganer et~al.(2002)Kaganer, Jenichen, Schippan, Braun, D\"aweritz, and
  Ploog}]{Kaganer2002}
\bibinfo{author}{V.~M. Kaganer}, \bibinfo{author}{B.~Jenichen},
  \bibinfo{author}{F.~Schippan}, \bibinfo{author}{W.~Braun},
  \bibinfo{author}{L.~D\"aweritz}, \bibinfo{author}{K.~H. Ploog},
\newblock \bibinfo{title}{Strain-mediated phase coexistence in {MnAs}
  heteroepitaxial films on {GaAs}: An x-ray diffraction study},
\newblock \bibinfo{journal}{Phys. Rev. B} \bibinfo{volume}{66}
  (\bibinfo{year}{2002}) \bibinfo{pages}{045305}.
\bibitem[{Breitwieser et~al.(2009)Breitwieser, Vidal, Graff, Marangolo,
  Eddrief, Boulliard, and Etgens}]{Breitwieser2009}
\bibinfo{author}{R.~Breitwieser}, \bibinfo{author}{F.~Vidal},
  \bibinfo{author}{I.~L. Graff}, \bibinfo{author}{M.~Marangolo},
  \bibinfo{author}{M.~Eddrief}, \bibinfo{author}{J.~C. Boulliard},
  \bibinfo{author}{V.~H. Etgens},
\newblock \bibinfo{title}{Phase transition and surface morphology of
  {MnAs}/{GaAs}(001) studied with in situ variable-temperature scanning
  tunneling microscopy},
\newblock \bibinfo{journal}{Phys. Rev. B} \bibinfo{volume}{80}
  (\bibinfo{year}{2009}) \bibinfo{pages}{045403}.
\bibitem[{Gumberidze et~al.(2010)Gumberidze, Trassinelli, Adrouche, Szabo,
  Indelicato, Haranger, Isac, Lamour, Le~Bigot, Merot, Prigent, Rozet, and
  Vernhet}]{Gumberidze2010}
\bibinfo{author}{A.~Gumberidze}, \bibinfo{author}{M.~Trassinelli},
  \bibinfo{author}{N.~Adrouche}, \bibinfo{author}{C.~I. Szabo},
  \bibinfo{author}{P.~Indelicato}, \bibinfo{author}{F.~Haranger},
  \bibinfo{author}{J.~M. Isac}, \bibinfo{author}{E.~Lamour},
  \bibinfo{author}{E.~O. Le~Bigot}, \bibinfo{author}{J.~Merot},
  \bibinfo{author}{C.~Prigent}, \bibinfo{author}{J.~P. Rozet},
  \bibinfo{author}{D.~Vernhet},
\newblock \bibinfo{title}{Electronic temperatures, densities, and plasma x-ray
  emission of a 14.5 {GHz} electron-cyclotron resonance ion source},
\newblock \bibinfo{journal}{Rev. Sci. Instrum.} \bibinfo{volume}{81}
  (\bibinfo{year}{2010}) \bibinfo{pages}{033303--10}.
\bibitem[{Ziegler et~al.(2008)Ziegler, Biersack, and Ziegler}]{Ziegler2008}
\bibinfo{author}{J.~F. Ziegler}, \bibinfo{author}{J.~P. Biersack},
  \bibinfo{author}{M.~D. Ziegler}, \bibinfo{title}{Stopping and Range of Ions
  in Matter}, \bibinfo{publisher}{SRIM Company}, \bibinfo{year}{2008}.

\end{thebibliography}







\end{document}